\documentclass{jnmp}

\usepackage{amsmath}

\JNMPnumberwithin{equation}{section}

\newtheorem{theorem}{Theorem}
\newtheorem{lemma}{Lemma}

\theoremstyle{definition}
\newtheorem*{example}{Example}

\begin{document}

\renewcommand{\evenhead}{M H Lee}
\renewcommand{\oddhead}{Tau Functions Associated to Pseudodifferential
Operators of Several Variables}

\thispagestyle{empty}

\FirstPageHead{9}{4}{2002}{\pageref{lee-firstpage}--\pageref{lee-lastpage}}{Article}

\copyrightnote{2002}{M H Lee}

\Name{Tau Functions Associated to Pseudodifferential
Operators of Several Variables}
\label{lee-firstpage}

\Author{Min Ho LEE}

\Address{Department of Mathematics, University of Northern Iowa,
Cedar Falls, IA 50614, U.S.A.\\
E-mail: lee@math.uni.edu}

\Date{Received April 26, 2002; Accepted June 3, 2002}

\begin{abstract}
\noindent
Pseudodifferential operators of several variables are formal Laurent series
in the formal inverses of $\partial_1, \ldots, \partial_n$ with $\partial_i = d/dx_i$ for
$1 \leq i \leq n$.  As in the single variable case, Lax equations can be
constructed using such pseudodifferential operators, whose solutions can be
provided by Baker functions.  We extend the usual notion of tau functions
to the case of pseudodifferential operators of several variables so that
each Baker function can be expressed in terms of the corresponding tau
function.
\end{abstract}

\section{Introduction}

One of the most actively studied areas in mathematics for the past few
decades is the theory of integrable nonlinear partial differential
equations (see e.g.~\cite{Ca91,Ch96,Di91,Ku00}).
Such equations are also known as soliton equations because they possess
localized nonlinear waves called solitons as solutions.  Examples of
soliton equations include many well-known equations in mathematical physics
such as the nonlinear Schr\"odinger equation, the Sine-Gordon equation, the
Korteweg-de Vries (KdV) equation, and the Katomtsev--Petviashvili (KP)
equation.

The main tool used in a systematic study of soliton equations is the notion
of Lax equations, which describe certain compatibility conditions for pairs
of differential operators.  A system of soliton equations called a KP
hierarchy is produced by a set of Lax equations, and as a result, solutions
of Lax equations can be used to construct solutions of the associated
soliton equations.  The interpretation of soliton equations in terms of Lax
equations leads to the derivation of the integrability as well as other
interesting properties of soliton equations.

A few decades ago, Krichever (see e.g.~\cite{Kr77}) introduced the method
of constructing an infinite dimensional subspace of ${\mathbb C} (( z))$
associated to some algebro-geometric data, where ${\mathbb C}(( z))$ is the
space of Laurent series.  This construction is nowadays called the
Krichever map, and it has been used successfully in the soliton theory and
is closely linked to the theory of moduli of algebraic curves (cf.~\cite{AD88,Kr77,SW85}).
 Thus the Krichever map provides a
connection of soliton theory with algebraic geometry, which is one of the
most intriguing features of the theory of soliton equations.  More
specifically, to each subspace of ${\mathbb C}(( z))$ produced by the Krichever
map there corresponds a so-called Baker--Akhiezer function, which determines
an algebro-geometric solution of a soliton equation (see~\cite{BB94,Ch96,Kr77,SW85}).
 Baker functions are a generalized
version of Baker--Akhiezer functions, and they supply formal solutions of
Lax equations.  Tau functions also play an important role in
algebro-geometric theory of solitons, and in particular, each Baker
function can be expressed in terms of the associated tau function.  Such an
expression of a Baker function in terms of a tau function is an important
contribution of the Japanese school (see e.g.~\cite{DK83}).  Tau functions
can be used to construct soliton solutions of soliton equations, and they
are essential in linking soliton theory to quantum field theory as well as
to the theory of Virasoro algebras or vertex operators.

Pseudodifferential operators are formal Laurent series in the formal
inverse $\partial^{-1}$ of the differentiation operator $\partial = d/dx$ with
respect to the single variable $x$, and they are essential ingredients in
the construction of Lax equations.  For this reason pseudodifferential
operators have played a major role in the theory of soliton equations.  In
a recent paper, Parshin~\cite{Pa99} studied pseudodifferential operators of
several variables by considering formal Laurent series in the formal
inverses of $\partial_1, \ldots, \partial_n$ with $\partial_i = d/dx_i$ for $1 \leq i \leq
n$.  Among other things, he constructed Lax equations associated to such
pseudodifferential operators and studied some of their properties.  Since
then, algebro-geometric connections of those pseudodifferential operators
have been studied in~\cite{Pa99a} and~\cite{Os01}, where the possibility of
extending the Krichever map to the case of higher dimensional varieties was
discussed.  Baker functions which provide solutions to Lax equations of
Parshin type have also been investigated in~\cite{L0f}, where some of the
properties of the usual Baker functions were extended to the case of
pseudodifferential operators of several variables.  The goal of this paper
is to prove the existence of tau functions associated to Baker functions
constructed in~\cite{L0f}.

\section{Pseudodifferential operators} \label{S:ps}

In this section we review pseudodifferential operators of several variables
studied by Parshin~\cite{Pa99} as well as the associated Lax equations.  We
also describe an example of a system of partial differential equations
determined by such a Lax equation.

We fix a positive integer $n$ and consider the variables $x_1, \ldots,
x_n$.  We denote by
\[
{\mathbb C} ((x_1)) \cdots ((x_n))
\]
the associated field of iterated Laurent series over $\mathbb C$, and let $P$
be the space of iterated formal Laurent series of the form
\[
P= {\mathbb C} ((x_1)) \cdots ((x_n))
\left(\left(\partial^{-1}_1\right)\right) \cdots
\left(\left(\partial^{-1}_n\right)\right)
\] in the formal inverses of the differential operators
\[
\partial_1 = \frac{\partial} {\partial x_1}, \ \ \ldots, \ \ \partial_n = \frac{\partial} {\partial x_n} .
\]
Throughout this paper we shall often use the usual multi-index notation.
Thus, given $\alpha = (\alpha_1, \ldots, \alpha_n) \in {\mathbb Z}^n$, we may write
\[
\partial^\alpha = \partial_1^{\alpha_1} \cdots \partial_n^{\alpha_n}, \qquad |\alpha| = \alpha_1 +
\cdots +\alpha_n ,
\]
with $\partial = (\partial_1, \ldots, \partial_n)$.  We also write $\alpha \geq \beta$ for
$\beta = (\beta_1, \ldots, \beta_n) \in {\mathbb Z}^n$ if $\alpha_i \geq \beta_i$ for
each~$i$, and use $\mathbf 0$ and $\mathbf 1$ to denote the elements $(0, \ldots, 0)$
and $(1, \ldots, 1)$ in ${\mathbb Z}^n$, respectively.  Thus, for example, an
element $\psi \in P$ can be written in the form
\begin{equation} \label{E:zx}
\psi = \sum_{\alpha \leq \nu} f_\alpha (x) \partial^\alpha
\end{equation}
for some $\nu \in {\mathbb Z}^n$.  We introduce a multiplication operation on
$P$ defined by the Leibniz rule, which means that
\[
\left( \sum_{\alpha} f_\alpha (x) \partial^\alpha \right) \left( \sum_{\beta} h_\beta (x)
\partial^\beta \right) = \sum_{\alpha, \beta} \sum_{\gamma \geq {\mathbf 0}}
\left(\begin{array}{c} \alpha \\ \gamma\end{array}\right) f_\alpha
(x) (\partial^\gamma h_\beta (x)) \partial^{\alpha+\beta -\gamma} ,
\]
where $\left(\begin{array}{c} \alpha \\ \gamma\end{array}\right) =
\left(\begin{array}{c} {\alpha_1}\\ {\gamma_1}\end{array}\right) \cdots
\left(\begin{array}{c} {\alpha_n}\\ {\gamma_n}\end{array}\right)$
for elements $\alpha = (\alpha_1, \ldots, \alpha_n)$ and $\gamma = (\gamma_1, \ldots,
\gamma_n)$ of ${\mathbb Z}^n$ with $\gamma \geq {\mathbf 0}$.  We now set
\[
{\mathbb Z}^n_+ = \{ \alpha \in {\mathbb Z}^n \mid \alpha \geq {\mathbf 0}, \, |\alpha| \geq 1 \},
\]
and assume that each coefficient $f_\alpha (x)$ in \eqref{E:zx} is a function
of the infinitely many variables $\{t_\alpha \mid \alpha \in {\mathbb Z}^n_+ \}$.  Let
${\mathbf e}_1 = (1,0,\ldots, 0)$, $\ldots$, ${\mathbf e}_n = (0,\ldots, 0,1)$ be the
standard basis for the ${\mathbb Z}$-module ${\mathbb Z}^n$, and assume that
\begin{equation} \label{E:sa}
t_{{\mathbf e}_1} = x_1, \ \ \ldots, \ \ t_{{\mathbf e}_n} = x_n .
\end{equation}
Thus we may write $\psi \in P$ in \eqref{E:zx} in the form
\[
\psi = \sum_{\alpha \leq \nu} f_\alpha (t) \partial^\alpha
\]
with $t = (t_\alpha)_{\alpha \in {\mathbb Z}_+^n}$.

If $\psi$ is an element of $P$ which can be written in the form
\[
\psi= \sum_{i= -\infty}^{\nu_n} a_i \partial_n^i = \sum_{i= -\infty}^{\nu_n}
a_i (t; \partial_1, \ldots, \partial_{n-1}) \partial_n^i
\]
with $\nu_n \geq 0$, we set
\[
\psi_+ = \sum_{i= 0}^{\nu_n} a_i \partial_n^i, \qquad \psi_- = \psi- \psi_+  =
\sum_{i= -\infty}^{-1} a_i \partial_n^i ;
\]
if $\nu_n < 0$, we set $\psi_+ = 0$ and $\psi_- = \psi$ .  Thus we
have $\psi = \psi_+ + \psi_-$ for all $\psi \in P$, and therefore $P$ can
be decomposed as
\[
P = P_+ + P_- ,
\]
where $P_+$ is the set of elements of $P$ of the form $\sum\limits_{i= 0}^m a_i
\partial_n^i$ for some nonnegative integer $m$, and $P_-$ is the set of
elements of the form $\sum\limits_{j= 0}^k b_j \partial_n^j$ with $k <0$.  Let $P^n$ be
the Cartesian product of $n$ copies of $P$, and consider an element $L =
(L_1, \ldots, L_n) \in P^n$ which satisfies the generalized Lax equation
\begin{equation} \label{E:4q}
\partial_{t_\alpha} L= [L^\alpha_+,L] = L^\alpha_+ L - L L^\alpha_+
\end{equation}
for all $\alpha \in {\mathbb Z}^n_+$, where $L^\alpha_+ = (L^\alpha)_+ = (L^{\alpha_1}_1
\cdots L^{\alpha_n}_n)_+ \in P_+$ and
\[
\partial_{t_\alpha} L= \frac{\partial L} {\partial t_\alpha} = \left( \frac {\partial L_1} {\partial
t_\alpha}, \ldots , \frac {\partial L_n} {\partial t_\alpha} \right) .
\]
Thus \eqref{E:4q} is equivalent to the system of equations
\[
\frac {\partial L_i} {\partial t_\alpha} = [L^\alpha_+,L_i]
\]
for $1\leq i \leq n$.

We now consider an element $\phi \in 1 + P_-$ satisfying the relation
\begin{equation} \label{E:w4}
\partial_{t_\alpha} \phi = - \left(\phi \partial^\alpha \phi^{-1}\right)_- \phi
\end{equation}
for each $\alpha \in {\mathbb Z}^n_+$, and set
\begin{equation} \label{E:ak}
L= \phi \partial \phi^{-1} = \left(\phi \partial_1 \phi^{-1}, \ldots, \phi \partial_n
\phi^{-1}\right) \in P^n .
\end{equation}
Thus, if $L = (L_1, \ldots, L_n)$, then each $L_i$ is of the form
\[ L_i = \phi \partial_i \phi^{-1} = \partial_i +u_i \]
for some $u_i \in P_-$.  Then it can be shown that the pseudodifferential
operator $L$ given by~\eqref{E:ak} satisfies the Lax equation~\eqref{E:4q}
for each $\alpha \in {\mathbb Z}^n_+$.  The Lax equation~\eqref{E:4q} also implies
the relation
\begin{equation} \label{E:p9}
\frac {\partial L^\beta_+} {\partial t_\alpha} - \frac {\partial L^\alpha_+} {\partial t_\beta} =
[ L^\alpha_+,  L^\beta_+]
\end{equation}
for all $\alpha, \beta \in {\mathbb Z}^n_+$ (see~\cite[Proposition 4]{Pa99}).  For
each pair $(\alpha, \beta)$ of elements of ${\mathbb Z}^n_+$ the
relation~\eqref{E:p9} determines a system of partial differential equations as can
be seen in the following example.

\begin{example}
We shall derive partial differential equations which are determined by the
Lax equation for $n=2$ associated to the pseudodifferential operators $L_1,
L_2 \in P$ given by
\begin{gather}
L_1 = \partial_2 + a \partial_1 \partial_2^{-1} + b \partial_2^{-2} + O\left(\partial_2^{-3}\right),
\label{E:l1}\\
L_2 = \partial_2 + c \partial_2^{-1} + d \partial_1 \partial_2^{-2} + O\left(\partial_2^{-3}\right)
\label{E:l2}
\end{gather}
for some functions $a = a (t)$, $b = b (t)$, $c = b (t)$ and $d = d (t)$
with $t = (t_\alpha)_{\alpha \in {\mathbb Z}_+^n}$.  We also consider the indices
\[
\alpha = (1,1), \qquad \beta = (1,2) ,
\]
so that $L^\alpha = L_1 L_2$ and $L^\beta = L_1 L_2^2$, where $L= (L_1, L_2) \in
P^2$.  Then by \eqref{E:p9}
the differential operators $L^\alpha_+$ and $L^\beta_+$ satisfy
\begin{equation} \label{E:pf}
\frac {\partial L^\beta_+} {\partial t_\alpha} - \frac {\partial L^\alpha_+} {\partial t_\beta} =
L^\alpha_+ L^\beta_+ - L^\beta_+ L^\alpha_+ .
\end{equation}
Using~\eqref{E:l1} and~\eqref{E:l2}, we obtain
\begin{gather*}
L_2^2 = \partial_2^2 + c_y \partial_2^{-1} + 2c + 2d \partial_1 \partial_2^{-1} + d_y \partial_1
\partial_2^{-2} + c^2 \partial_2^{-2} + O\left(\partial_2^{-3}\right),\\
L_1 L_2 = \partial_2^2 + a\partial_1 + c + O\left(\partial_2^{-1}\right) ,\\
L_1 L_2^2 = \partial_2^3 +a \partial_1 \partial_2 + 2c \partial_2 + 2d \partial_1 + 3c_y+ b +
O\left(\partial_2^{-1}\right) ,
\end{gather*}
where the subscripts $x$ and $y$ denote the partial derivatives with
respect to $x= x_1$ and $y = x_2$, respectively.  Hence we have
\begin{gather}
L^\alpha_+ = (L_1 L_2)_+ = \partial_2^2 + a\partial_1 + c,
\label{E:w1}\\
L^\beta_+ = \left(L_1 L_2^2\right)_+ = \partial_2^3 +a \partial_1 \partial_2 + 2c \partial_2 + 2d \partial_1 +
3c_y + b .
\label{E:w2}
\end{gather}
Using~\eqref{E:w1} and~\eqref{E:w2}, we obtain
\begin{gather*}
L^\alpha_+ L^\beta_+ = \partial_2^5 + 2a \partial_1 \partial_2^3 + 3c \partial_2^3 + (2a_y +2d) a
\partial_1 \partial_2^2 + (7 c_y +b) \partial_2^2 + a^2 \partial_1^2 \partial_2\\
\phantom{L^\alpha_+ L^\beta_+ =}{} + (a_{yy} + 4d_y + aa_x + 3ac)
\partial_1 \partial_2 + \left(8c_{yy} +2b_y+ 2ac_x + 2c^2\right) \partial_2\\
\phantom{L^\alpha_+ L^\beta_+ =}{}+ 2ad \partial_1^2 + (2d_{yy} +2ad_x + 3ac_y +ab +cd) \partial_1\\
\phantom{L^\alpha_+ L^\beta_+ =}{} + 3c_{yyy} + b_{yy} + 3ac_{yx} + ab_x + 3cc_y +bc,\\
L^\beta_+ L^\alpha_+ = \partial_2^5 + 2a \partial_1 \partial_2^3 + 3c \partial_2^3 + (3a_y +2d) a
\partial_1 \partial_2^2 + (6 c_y +b) \partial_2^2 + a^2 \partial_1^2 \partial_2\\
\phantom{L^\beta_+ L^\alpha_+ =}{} + (3a_{yy} + aa_x + 3ac) \partial_1 \partial_2 +
\left(3c_{yy} + ac_x +2c^2\right) \partial_2 + (aa_y +2ad) \partial_1^2\\
\phantom{L^\beta_+ L^\alpha_+ =}{} + (a_{yyy} +aa_{yx} +4a c_y +2 a_y c +2a_x d +ab +2cd)
\partial_1\\
\phantom{L^\beta_+ L^\alpha_+ =}{} + c_{yyy} + ac_{yx} + 5cc_y +2c_x d +bc.
\end{gather*}
If we set $t_\alpha =s$ and $t_\beta =t$, then by \eqref{E:w1} and \eqref{E:w2}
the left hand side of~\eqref{E:pf} becomes
\[ \frac {\partial L^\beta_+} {\partial s} - \frac {\partial L^\alpha_+} {\partial t} = a_s \partial_1
\partial_2 + 2c_s \partial_2 + (2d_s -a_t) \partial_1 + 3 c_{ys} + b_s - c_t .
\]
Thus by comparing the coefficients we see that~\eqref{E:pf} determines the
system of partial differential equations given by
\begin{gather*}
a_y = c_y =0,\qquad
2c_s = 2b_y + ac_x,\\
a_t + 2ad_x +2d_{yy} = 2a_x d +2d_s + cd,\qquad
b_s +2c_x d = ab_x +b_{yy} + c_t.
\end{gather*}
\end{example}

\section{Baker functions} \label{S:bf}

Baker functions associated to pseudodifferential operators of several
variables discussed in Section \ref{S:ps} were introduced in \cite{L0f}.
As in the single variable case, these Baker functions provide solutions of
Lax equations of the from \eqref{E:4q}.  In this section we review the
construction of such Baker functions.

First, we need to introduce an additional set of complex variables $z_1,
\ldots, z_n$.  We then consider the formal series given by
\begin{equation} \label{E:7n}
\xi (t, z) = \sum_{\alpha \in {\mathbb Z}^n_+} t_\alpha z^\alpha ,
\end{equation}
where $z = (z_1, \ldots, z_n)$ so that $z^\alpha = z_1^{\alpha_1} \cdots
z_n^{\alpha_n}$ for $\alpha = (\alpha_1, \ldots, \alpha_n)$.  If $\phi \in 1+ P_-$ is
as in Section~\ref{S:ps} satisfying~\eqref{E:w4}, we define the associated
{\it Baker function\/} $w$ by
\begin{equation} \label{E:q0}
w = w(t, z) = \phi e^{\xi (t, z)} .
\end{equation}
Since $x_i = t_{{\mathbf e}_i}$ for $1 \leq i \leq n$ by~\eqref{E:sa}, we see that
\[
 \partial_i e^{\xi (t, z)} = \frac {\partial} {\partial x_i} e^{\xi (t, z)} = \frac
{\partial} {\partial t_{\mathbf e_i}} e^{\xi (t, z)} = z^{{\mathbf e}_i} e^{\xi (t, z)} = z_i e^{\xi
(t, z)}.
\]
Thus, if $\alpha = (\alpha_1, \ldots, \alpha_n) \in {\mathbb Z}^n_+$, we have
\begin{gather*}
\partial^\alpha e^{\xi (t, z)} = \partial_1^{\alpha_1} \cdots \partial_n^{\alpha_n} e^{\xi (t, z)}
= \partial_{\mathbf e_1}^{\alpha_1} \cdots \partial_{\mathbf e_n}^{\alpha_n} e^{\xi (t, z)}\\
\phantom{\partial^\alpha e^{\xi (t, z)}}{}=
z_1^{\alpha_1} \cdots z_n^{\alpha_n} e^{\xi (t, z)} = z^\alpha e^{\xi (t, z)}.
\end{gather*}
Hence, if $\phi = 1 + \sum\limits_\alpha a_\alpha (t) \partial^\alpha\in 1+ P_-$, then the Baker
function in~\eqref{E:q0} can be written in the form
\begin{equation} \label{E:ne}
w (t,z) = \widehat{w} (t,z) e^{\xi (t, z)} ,
\end{equation}
where $\widehat{w} (t,z)$ is a formal power series in $z_1, \ldots, z_n$ of the
form
\[
 \widehat{w} (t,z) = 1 + \sum_\alpha a_\alpha (t) z^\alpha .
\]
If $L= (L_1, \ldots, L_n) = \phi \partial \phi^{-1} \in P^n$ is an element
associated to $\phi \in 1+ P_-$ satisfying~ \eqref{E:w4} as in~\eqref{E:ak},
then the Baker function $w$ given by~\eqref{E:q0} satisfies $Lw = zw$, that
is, $L_i w = z_i w$ for each $i$ (see~\cite[Lemma 3.1]{L0f}).  In addition,
it can also be shown that $\partial_{t_\alpha} w = L^\alpha_+ w$ for each $\alpha \in
{\mathbb Z}^n_+$ (cf.~\cite[Lemma 3.2]{L0f}).

Given an element $\psi = \sum\limits_{\alpha \leq \nu} f_\alpha (t) \partial^\alpha \in P$, we
define its adjoint $\psi^* \in P$ by
\begin{equation} \label{E:yb}
\psi^* = \sum_{\alpha \leq \nu} (-1)^{|\alpha|} \partial^\alpha f_\alpha (t) ,
\end{equation}
and its residue with respect to $\partial$ by
\[
{\rm Res}_\partial \psi = f_{-{\mathbf 1}} (t) = f_{(-1, \ldots, -1)} (t) .
\]
On the other hand, if $h (z) = h (z_1, \ldots, z_n)$ is a Laurent series in
$z_1, \ldots, z_n$ which can be written in the form $h(z) = \sum\limits_\alpha b_\alpha
z^\alpha$, then its residue with respect to $z$ is given by
\begin{equation} \label{E:h6}
{\rm Res}_z h(z) = b_{-\mathbf 1} = b_{(-1, \ldots, -1)} .
\end{equation}
If $\psi = \sum\limits_\alpha a_\alpha
\partial^\alpha \in P$ and $\eta = \sum\limits_\beta b_\beta \partial^\beta
\in 1 +P_-$, then we have
\[
{\rm Res}_z \left(\psi e^{\xi (t,z)}\right) \left(\eta e^{-\xi (t,z)}\right) = {\rm Res}_\partial \psi \eta^*,
\]
where $\eta^*$ is the adjoint of $\eta$ given by~\eqref{E:yb} (see~\cite[Lemma 3.3]{L0f}).

We define the adjoint $w^*$ of the Baker function $w$ in~\eqref{E:q0} by
\begin{equation} \label{E:r4}
w^* (t,z) = (\phi^*)^{-1} e^{-\xi (t,z)} ,
\end{equation}
where $\phi^*$ is the adjoint of $\phi$ given by \eqref{E:yb}.  Then it can
be shown that the Baker function~$w$ in~\eqref{E:q0} satisfies
\begin{equation} \label{E:w9}
{\rm Res}_z w (t',z) w^\ast (t, z) = 0
\end{equation}
for all $t$, $t'$ (see~\cite{L0f}).

We now consider the subset $\widehat{P}_-$ of $P_-$ defined by
\begin{equation} \label{E:h5}
\widehat{P}_- = \left\{ \sum_\alpha f_\alpha (t) \partial^\alpha \; \bigg| \; \alpha \leq
-{\mathbf 1} = (-1, \ldots, -1) \ \mbox{whenever} \ f_\alpha (t) \neq 0 \right\} .
\end{equation}
Then the following theorem extends the result in~\cite[Proposition~7.3.5]{Di91}
to the case of several variables.

\begin{theorem}
Let $w$ and $w^\#$ be formal power series of the form
\[
w = \phi e^{\xi (t, z)}, \qquad w^\# = \psi e^{-\xi (t, z)}
\]
with $\phi, \psi \in 1+ \widehat{P}_-$ satisfying the condition
\[
{\rm Res}_z \left(\partial^\alpha w w^\#\right) = 0.
\]
Then there exists an operator $L= (L_1, \ldots, L_n) \in P^n$ with $L_i =
\partial_i + u_i$ and $u_i \in P_-$ for $1\leq i \leq n$ which satisfies the Lax
equation~\eqref{E:4q} with $w$ and $w^\#$ being the associated Baker
function and adjoint Baker function, respectively.
\end{theorem}

\begin{proof}
See~\cite[Theorem 3.6]{L0f}.
\end{proof}

\section{Tau functions}

In this section we extend the notion of tau functions associated
to the usual pseudodifferential operators to the case of
pseudodifferential operators of several variables.  As in the
single variable case, a Baker function given by~\eqref{E:q0} can
be expressed in terms of such a~tau function.

Let $t= (t_\alpha)_{\alpha \in {\mathbb Z}_+^n}$ and $z = (z_1, \ldots, z_n)$ be the
complex variables considered in Section~\ref{S:bf}.  Given a vector $s =
(s_1, \ldots, s_n) \in {\mathbb C}^n$, we define the operator $G(s)$ on
functions of the form $f(t,z) = f((t_\alpha)_{\alpha \in {\mathbb Z}_+^n},(z_1,
\ldots, z_n))$ by
\begin{equation} \label{E:yr}
G(s)  f(t,z) = f \left(\left(t_\alpha - \alpha^{-1} s^{-\alpha}\right)_{\alpha \in {\mathbb Z}_+^n}, z\right),
\end{equation}
where $\alpha^{-1} s^{-\alpha} = \alpha_1^{-1} \cdots \alpha_n^{-1} s_1^{-\alpha_1} \cdots
s_n^{-\alpha_n}$ according to the multi-index notation.  Thus, if $\xi (t,z)$ is
as in~\eqref{E:7n}, we have
\[
G(s)  \xi (t,z) = \sum_{\alpha \in {\mathbb Z}_+^n} \left(t_\alpha - \alpha^{-1}
s^{-\alpha}\right) z^\alpha = \xi (t,z) - \sum_{\alpha \in {\mathbb Z}_+^n} \alpha^{-1} s^{-\alpha}
z^\alpha .
\]
Hence it follows that $G(s)$ operates on the Baker function $w(t,z)$
in~\eqref{E:q0} associated to an element $\phi \in 1+P_-$ and on the adjoint
Baker function $w(t,z)$ in~\eqref{E:r4} by
\begin{gather*}
G(s)  w(t,z) = w(t,z) \exp \left( - \sum_{\alpha \in {\mathbb Z}_+^n}
\alpha^{-1} s^{-\alpha} z^\alpha \right) ,\\
G(s)  w^* (t,z) = w^* (t,z) \exp \left( \sum_{\alpha \in {\mathbb Z}_+^n}
\alpha^{-1} s^{-\alpha} z^\alpha \right) .
\end{gather*}
Using the relation
\[
\ln \left( 1- \sum^n_{r=1} \frac {z_r} {s_r} \right) = - \sum_{\alpha \in
{\mathbb Z}_+^n} \frac {z_1^{\alpha_1} \cdots z_n^{\alpha_n}} {\alpha_1 \cdots \alpha_n
s_1^{\alpha_1} \cdots s_n^{\alpha_n}} = - \sum_{\alpha \in {\mathbb Z}_+^n} \alpha^{-1}
s^{-\alpha} z^\alpha,
\]
we see that the operation of $G(s)$ on $w^* (t,z)$ can be written in the
form
\begin{equation} \label{E:e9}
G(s) w^* (t,z) = w^* (t,z) \left( 1- \sum^n_{r=1} \frac {z_r} {s_r}
\right)^{-1} .
\end{equation}
We now consider some calculations involving the residue operator ${\rm Res}_z$
given by~\eqref{E:h6}.

\begin{lemma} \label{L:u1}
Let $s = (s_1, \ldots, s_n) \in {\mathbb C}^n$, and consider a formal power
series of the form $\eta (z) = 1+ \sum\limits_{\alpha \leq -{\mathbf 1}} f_\alpha z^\alpha$.
Then we have
\[ {\rm Res}_z \eta (z) \left( 1- \sum^n_{r=1} \frac {z_r} {s_r} \right)^{-1} =
s_1 \cdots s_n (\eta (s) -1) \]
for all $z = (z_1, \ldots, z_n) \in {\mathbb C}^n$.
\end{lemma}

\begin{proof}
First, we write the formal power series $\eta (z)$ in the form
\[
\eta (z) = 1 + \sum^{-1}_{\alpha_1 = -\infty} \cdots \sum^{-1}_{\alpha_n =
  -\infty} f_{(\alpha_1, \ldots, \alpha_n)} z_1^{\alpha_1} \cdots z_n^{\alpha_n} .
\]
Using this and the power series expansion
\begin{gather*}
\left( 1- \sum^n_{r=1} \frac {z_r} {s_r} \right)^{-1} = \sum^\infty_{r=0}
\left(s_1^{-1} z_1 + \cdots + s_n^{-1} z_n\right)^r
= \sum_{\beta \geq {\mathbf 0}} s^{-\beta} z^\beta = \sum_{\beta \geq {\mathbf 0}} \frac
{z_1^{\beta_1} \cdots z_n^{\beta_n}} {s_1^{\beta_1} \cdots s_n^{\beta_n}}
\end{gather*}
with $\beta = (\beta_1, \ldots, \beta_n)$, we see that
\begin{gather*}
{\rm Res}_z \eta (z) \left( 1- \sum^n_{r=1} \frac {z_r} {s_r} \right)^{-1} =
  \sum^{-1}_{\alpha_1 = -\infty} \cdots \sum^{-1}_{\alpha_1 = -\infty} \frac
  {f_{(\alpha_1, \ldots, \alpha_n)}} {s_1^{-\alpha_1-1} \cdots s_n^{-\alpha_n-1}}\\
\qquad{}= s_1 \cdots s_n \sum^{-1}_{\alpha_1 = -\infty} \cdots \sum^{-1}_{\alpha_n =
  -\infty} f_{(\alpha_1, \ldots, \alpha_n)} s_1^{\alpha_1} \cdots s_n^{\alpha_n}\\
\qquad{}= s_1 \cdots s_n \sum_{\alpha \leq -\mathbf 1} f_\alpha s^\alpha
= s_1 \cdots s_n (\eta (s) -1);
\end{gather*}
hence the lemma follows.
\end{proof}

\begin{lemma} \label{L:u2}
Let $s = (s_1, \ldots, s_n)$ and $s' = (s'_1, \ldots, s'_n)$ be elements of
${\mathbb C}^n$, and consider a formal power series of the form $\eta (z) = 1+
\sum\limits_{\alpha \leq - {\mathbf 1}} f_\alpha z^\alpha$.  Then we have
\begin{gather*}
{\rm Res}_z \eta (z) \left( 1- \sum^n_{r=1} \frac {z_r} {s_r} \right)^{-1}
\left( 1- \sum^n_{r=1} \frac {z_r} {s'_r} \right)^{-1} \\
\qquad {}= (\eta (s) - 1) \sum_{\alpha \geq {\mathbf 0}} \frac {s^{\alpha+1}} {{s'}^{\alpha}}
 = (\eta (s') - 1) \sum_{\alpha \geq {\mathbf 0}} \frac {{s'}^{\alpha+1}} {s^{\alpha}}
\end{gather*}
for all $z = (z_1, \ldots, z_n) \in {\mathbb C}^n$.
\end{lemma}

\begin{proof}
Using power series expansions and the formal relation
\[
 \sum^\infty_{r=0} (u_1 + \cdots +u_n)^r = \sum_{\alpha \geq {\mathbf 0}} u^\alpha
\]
for each $n$-tuple $u = (u_1, \ldots, u_n)$, we see that
\begin{gather*}
{\rm Res}_z \eta (z) \left( 1 - \sum^n_{r=1} \frac {z_r} {s_r} \right)^{-1}
\left( 1- \sum^n_{r=1} \frac {z_r} {s'_r} \right)^{-1}\\
\qquad {}= {\rm Res}_z \eta (z) \left( \sum^\infty_{r=0} \left( \sum^n_{r=1} \frac
{z_r} {s_r} \right)^r \right) \left( \sum^\infty_{r=0} \left(
\sum^n_{r=1} \frac {z_r} {{s'}_r} \right)^r \right)\\
\qquad {}= {\rm Res}_z \eta (z)
\left(  \sum_{\beta \geq {\mathbf 0}} \frac {z^\beta} {s^\beta} \right) \left(
\sum_{\gamma \geq {\mathbf 0}} \frac {z^\gamma} {{s'}^\gamma} \right)
= \sum_{\alpha \leq -{\mathbf 1}} f_\alpha  \sum_{\beta +\gamma = -\alpha-{\mathbf 1}}
\frac 1{s^\beta  {s'}^\gamma},
\end{gather*}
where the second summation in the previous line is over multi-indices $\beta,
\gamma \geq {\mathbf 0}$ such that $\beta + \gamma = - \alpha -{\mathbf 1}$.
Using $\beta = -\gamma -\alpha-1$, we obtain
\begin{gather*}
{\rm Res}_z \eta (z) \left( 1- \sum^n_{r=1} \frac {z_r} {s_r} \right)^{-1}
\left( 1- \sum^n_{r=1} \frac {z_r} {s'_r} \right)^{-1} \\
\qquad {}= \sum_{\alpha \leq
  -{\mathbf 1}} f_\alpha \sum_{\gamma \geq {\mathbf 0}} \frac 1{s^{-\gamma-\alpha-1} {s'}^\gamma}
= \sum_{\alpha \leq -{\mathbf 1}} f_\alpha  s^\alpha \sum_{\gamma \geq {\mathbf 0}} \frac
{s^{\gamma+1}} {{s'}^\gamma}
= (\eta (s)  - 1) \sum_{\gamma \geq {\mathbf 0}} \frac {s^{\gamma+1}} {{s'}^{\gamma}}.
\end{gather*}
Similarly, by using $\gamma = -\beta -\alpha-1$ we have
\begin{gather*}
{\rm Res}_z \eta (z) \left( 1- \sum^n_{r=1} \frac {z_r} {s_r} \right)^{-1}
\left( 1- \sum^n_{r=1} \frac {z_r} {s'_r} \right)^{-1} \\
\qquad {}= \sum_{\alpha \leq
  -{\mathbf 1}} f_\alpha  \sum_{\beta \geq {\mathbf 0}} \frac 1{s^\beta {s'}^{-\beta-\alpha -1}}
= \sum_{\alpha \leq -{\mathbf 1}} f_\alpha  {s'}^\alpha \sum_{\beta \geq {\mathbf 0}} \frac
{{s'}^{\beta+1}} {s^\beta}
= (\eta (s')  - 1) \sum_{\beta \geq {\mathbf 0}} \frac {{s'}^{\beta+1}} {s^{\beta}}.
\end{gather*}
Hence the lemma follows.
\end{proof}

We now state the main theorem in this section, which shows the existence of
the tau function $\tau (t)$ corresponding to a Baker function of the type
discussed in Section~\ref{S:bf}.

\begin{theorem} \label{T:hw}
Let $w(t,z)$ be the Baker function in~\eqref{E:q0} corresponding to an
element $\phi \in 1+P_-$, and let $\widehat{w} (t,z)$ be the associated formal
power series given by~\eqref{E:ne}.  Then there is a function $\tau (t)$
with $t = (t_\alpha)_{\alpha \in {\mathbb Z}_+^n}$ such that
\[
\widehat{w} (t,z) = G(z) \tau (t) / \tau (t)
\]
for $z \in {\mathbb C}^n$ and $t = (t_\alpha)_{\alpha \in {\mathbb Z}_+^n}$, where $G(z)$
is the operator given by~\eqref{E:yr}.
\end{theorem}

\begin{proof}
By \eqref{E:w9} we have
\[
{\rm Res}_z w(t,z) G(s) w^* (t,z) = 0
\]
for each $s \in {\mathbb C}^n$.  Using this and \eqref{E:e9}, we have
\[
{\rm Res}_z \widehat{w}(t,z) G(s) \widehat{w}^* (t,z) \left( 1- \sum^n_{r=1} \frac
{z_r} {s_r} \right)^{-1} = 0 .
\]
Thus by Lemma~\ref{L:u1} we see that
\[
s_1 \cdots s_n (\widehat{w}(t,s) G(s) \widehat{w}^* (t,s) -1) = 0 ;
\]
hence we obtain
\begin{equation} \label{E:r1}
\widehat{w}(t,s)^{-1} = G(s) \widehat{w}^* (t,s) .
\end{equation}
Similarly, we have
\[
{\rm Res}_z w(t,z) G(s) G(s') w^* (t,z) = 0
 \]
for all $s, s' \in {\mathbb C}^n$, which implies that
\[
{\rm Res}_z \widehat{w}(t,z) G(s) G(s') \widehat{w}^* (t,z) \left( 1- \sum^n_{r=1}
\frac {z_r} {s_r} \right)^{-1} \left( 1- \sum^n_{r=1} \frac {z_r} {s'_r}\right)^{-1} = 0 .
\]
Using this and applying Lemma~\ref{L:u2} to the formal power series
\[
\phi (z) = \widehat{w}(t,z) G(s) G(s') \widehat{w}^* (t,z) ,
\]
we obtain
\[
\widehat{w}(t,s) G(s) G(s') \widehat{w}^* (t,s) = \widehat{w}(t,s') G(s) G(s') \widehat{w}^*
(t,s') =1 .
\]
By combining this with~\eqref{E:r1} we have
\begin{equation} \label{E:mb}
\widehat{w}(t,s) (G(s') \widehat{w}(t,s))^{-1} = \widehat{w}(t,s) (G(s)
\widehat{w}(t,s'))^{-1} .
\end{equation}
We now set
\[
h(t,s) = \ln (\widehat{w}(t,s)) .
\]
Then by taking the logarithm of both sides of \eqref{E:mb} we obtain
\[
(1-G(s')) h(t,s) = (1-G(s)) h(t,s') .
\]
Replacing $s$ and $s'$ by $z$ and $\zeta$, respectively, gives us
\begin{equation} \label{E:r2}
h(t,z) - G(\zeta) h(t,z) = h(t,\zeta) - G(z) h(t,\zeta) .
\end{equation}
For each $k \in \{ 1, \ldots, n\}$ we define the differential operator
${\mathcal D}_k (z)$ by
\[
{\mathcal D}_k (z) = \sum_{\alpha \in {\mathbb Z}_+^n} \alpha_k \alpha^{-1} z^{-\alpha- {\mathbf
  e}_k} \partial_\alpha - \frac \partial {\partial z_k} ,
\]
where $\partial_\alpha
= \partial_{t_\alpha} = \partial/\partial t_\alpha$ with $\alpha = (\alpha_1, \ldots,
\alpha_n)$.  For any function $\varphi (t)$, we have
\begin{gather*}
{\mathcal D}_k (z) G(z) \varphi (t) = {\mathcal D}_k (z) \varphi \left(\left(t_\alpha - \alpha^{-1}
  z^{-\alpha}\right)_{\alpha \in {\mathbb Z}_+^n}\right)\\
\qquad {}= \sum_{\alpha \in {\mathbb Z}_+^n}
\alpha_k \alpha^{-1} z^{-\alpha- {\mathbf e}_k} \partial_\alpha \varphi
  \left(\left(t_\alpha - \alpha^{-1} z^{-\alpha}\right)_{\alpha \in \mathbb Z_+^n}\right)\\
\qquad {} - \sum_{\alpha \in {\mathbb Z}_+^n} \alpha_k \alpha^{-1} z^{-\alpha- {\mathbf e}_k}
  \partial_\alpha \varphi\left(\left(t_\alpha - \alpha^{-1} z^{-\alpha}
\right)_{\alpha \in {\mathbb Z}_+^n}\right) =0 .
\end{gather*}
Using this and~\eqref{E:r2}, we see that
\[
{\mathcal D}_k (z) h(t,z) - G(\zeta) {\mathcal D}_k (z) h(t,z) = {\mathcal D}_k (z) h(t,
\zeta) = \sum_{\alpha \in {\mathbb Z}_+^n}
\alpha_k \alpha^{-1} z^{-\alpha- {\mathbf e}_k} \partial_\alpha h(t,\zeta) .
\]
Thus for each $\beta \in {\mathbb Z}_+^n$ we obtain
\begin{gather*}
{\rm Res}_z z^\beta {\mathcal D}_k (z) h(t,z) - G(\zeta) \,{\rm Res}_z z^\beta {\mathcal D}_k (z)
h(t,z)\\
\qquad {}= {\rm Res}_z \sum_{\alpha \in {\mathbb Z}_+^n}
\alpha_k \alpha^{-1} z^{-\alpha+\beta - {\mathbf e}_k}
\partial_\alpha h(t,\zeta)
= \beta_k (\beta+{\mathbf 1}-{\mathbf e}_k)^{-1} \partial_{\beta+{\mathbf 1}-{\mathbf e}_k} h(t,\zeta) ,
\end{gather*}
where we used the fact that the $k$-component of $\beta+{\mathbf 1}-{\mathbf e}_k$ is
$\beta_k$.  Thus, if we set $a_{\alpha,k} = {\rm Res}_z z^\alpha \mathcal D_k (z) h(t,z)$
for each $\alpha \in {\mathbb Z}_+^n$, we have
\begin{equation} \label{E:r3}
(1- G(\zeta)) a_{\alpha,k} = \alpha_k (\alpha+{\mathbf 1}-{\mathbf e}_k)^{-1}
\partial_{\alpha+{\mathbf 1}-{\mathbf e}_k} h(t,\zeta)
\end{equation}
for all $\alpha \in {\mathbb Z}_+^n$ and $k \in \{ 1, \ldots, n\}$.  Hence we
obtain
\begin{gather*}
\alpha_k (\alpha+{\mathbf 1}-{\mathbf e}_k)^{-1}
 \partial_{\alpha+{\mathbf 1}-{\mathbf e}_k} a_{\beta,k} - \beta_k
(\beta+{\mathbf 1}-{\mathbf e}_k)^{-1} \partial_{\beta+{\mathbf 1}-{\mathbf e}_k} a_{\alpha,k}\\
\qquad {}= G(\zeta) \left(\alpha_k (\alpha+{\mathbf 1}-{\mathbf e}_k)^{-1}
\partial_{\alpha+{\mathbf 1}-{\mathbf e}_k} a_{\beta,k}
- \beta_k (\beta+{\mathbf 1}-{\mathbf e}_k)^{-1} \partial_{\beta+{\mathbf 1}-
{\mathbf e}_k} a_{\alpha,k}\right) ,
\end{gather*}
which implies that
\[
\alpha_k (\alpha+{\mathbf 1}-{\mathbf e}_k)^{-1} \partial_{\alpha+{\mathbf 1}-
{\mathbf e}_k} a_{\beta,k} = \beta_k
(\beta +{\mathbf 1}-{\mathbf e}_k)^{-1} \partial_{\beta+{\mathbf 1}-{\mathbf e}_k} a_{\alpha,k}
\]
for all $\alpha, \beta \in {\mathbb Z}_+^n$.  Therefore there is a function $\tau
(t)$ such that
\[
a_{\alpha,k} = -  \alpha_k (\alpha+{\mathbf 1}-{\mathbf e}_k)^{-1}
\partial_{\alpha+{\mathbf 1}-{\mathbf e}_k}
\ln \tau (t) .
\]
By combining this with \eqref{E:r3} we obtain
\begin{gather*}
\partial_{\alpha+{\mathbf 1}-{\mathbf e}_k} h(t,\zeta)
= \alpha_k^{-1} (\alpha+{\mathbf 1}-{\mathbf e}_k)
(1-G(\zeta)) a_{\alpha,k}
= - (1-G(\zeta)) \partial_{\alpha+{\mathbf 1}-{\mathbf e}_k} \ln \tau (t)
\end{gather*}
for all $\alpha \in {\mathbb Z}_+^n$ and $k \in \{ 1, \ldots, n\}$; hence we see
that
\[
h(t,\zeta) = - (1-G(\zeta)) \ln \tau (t) .
\]
Thus it follows that
\begin{gather*}
\ln (\widehat{w}(t,\zeta)) = h(t,\zeta) = -\ln \tau (t) + G(\zeta) \ln \tau (t)
= \ln (G(\zeta) \tau (t)/\tau (t)) .
\end{gather*}
Thus we obtain
\[
\widehat{w}(t,\zeta)= G(\zeta) \tau (t)/\tau (t) ,
\]
and therefore the proof of the theorem is complete.
\end{proof}

\section{Concluding remarks}

As is mentioned in the introduction, Baker functions associated to
single-variable pseudodifferential operators provide formal solutions of
soliton equations.  Baker functions for pseudodifferential operators of
several variables also determine solutions of soliton equations, and by
Theorem~\ref{T:hw} we see that the Baker function in~\eqref{E:q0} can be
written in the form
\begin{gather*}
w (t,z) = \widehat{w} (t,z) e^{\xi (t,z)} = (G(z) \tau (t)/\tau (t)) e^{\xi (t,z)} ,
\end{gather*}
where $\xi (t,z) = \sum\limits_{\alpha \in {\mathbb Z}_+^n} t_\alpha z^\alpha$.  The function
$\tau (t)$ with $t = (t_\alpha)_{\alpha \in {\mathbb Z}_+^n}$ is a tau function for
pseudodifferential operators of several variables.  Thus we have extended
the expression of a Baker function in term of the corresponding tau
function to the case of pseudodifferential operators of several variables.

\label{lee-lastpage}


\begin{thebibliography}{99}
\small

\bibitem{AD88}
Arbarello E, De Concini C, Kac V~G and Procesi~C, Moduli
  Spaces of Curves and Representation Theory,
{\it Comm. Math. Phys.} {\bf 117} (1988), 1--36.

\bibitem{BB94}
Belokolos E, Bobenko A, Enol'skii V, Its A and Matveev V,
  Algebro-Geometric Approach to Nonlinear Integrable Equations,
  Springer-Verlag, Heidelberg, 1994.

\bibitem{Ca91}
Carroll R, Topics in Soliton Theory, North-Holland, Amsterdam,
1991.

\bibitem{Ch96}
Cherednik I, Basic Methods of Soliton Theory, World Scientific,
  Singapore, 1996.

\bibitem{DK83}
Date E, Kashiwara M, Jimbo M and Miwa T, Transformation Groups
  for Soliton Equations, in Nonlinear Integrable Systems~--- Classcal Theory
and Quantum Theory (Kyoto, 1981), 39--119, World Scientific, Singapore, 1983.

\bibitem{Di91}
Dickey L, Soliton Equations and Hamiltonian Systems, World
  Scientific, Singapore, 1991.

\bibitem{Kr77}
Krichever A, Methods of Algebraic Geometry in the Theory of
  Non-Linear Equations, {\it Russian Math. Surveys} {\bf 32} (1977), 185--213.

\bibitem{Ku00}
Kupershmidt B, KP or mKP: Noncommutative Mathematics of
  Lagrangian, Hamiltonian, and Integrable Systems, Amer. Math. Soc.,
  Providence, 2000.

\bibitem{L0f}
Lee M~H, Pseudodifferential Operators of Several Variables and
Baker Functions, {\it Lett. Math. Phys.}, {\bf 60} (2002), 1--8.

\bibitem{Os01}
Osipov D, Krichever Correpondence for Algebraic Varieties,
{\it Izvestiya RAN: Ser. Math.} {\bf 65} (2001), 91--128.

\bibitem{Pa99a}
Parshin A, The Krichever Correspondence for Algebraic Surfaces, {\it Funct. Anal. Appl.} {\bf 35} (2001), 74--76.

\bibitem{Pa99}
Parshin A, On a Ring of Formal Pseudo-Differential Operators, {\it Proc.
  Steklov Inst. Math.} {\bf 224} (1999), 266--280.

\bibitem{SW85}
Segal G and Wilson G, Loop Groups and Equations of KdV, {\it Publ.
  Math. I.H.E.S.} {\bf 61} (1985), 5--65.

\end{thebibliography}
\end{document}